\documentclass[prc,preprint,showpacs,amsmath,amssymb]{revtex4}
\usepackage{graphicx}
\usepackage{dcolumn}
\usepackage{bm}
\usepackage{booktabs}
\usepackage{color}
\begin{document}

\title{Characterizing the astrophysical S-factor for $^{12}$C+$^{12}$C with 
wave-packet dynamics}

\author{Alexis Diaz-Torres$^1$ and Michael Wiescher$^{2}$}

\affiliation{$^1$ Department of Physics, University of Surrey, 
Guildford GU2 7XH, UK \\
$^{2}$JINA and University of Notre Dame, Indiana 46656, USA}

\date{\today}

\begin{abstract}
A quantitative study of the astrophysically important sub-barrier fusion of $^{12}$C+$^{12}$C is presented. Low-energy collisions are described in the body-fixed reference frame using wave-packet dynamics within a nuclear molecular picture. A collective Hamiltonian drives the time propagation of the wave-packet through the collective potential-energy landscape. The fusion imaginary potential for specific dinuclear configurations is crucial for understanding the appearance of resonances in the fusion cross section. The theoretical sub-barrier fusion cross sections explain some observed resonant structures in the astrophysical S-factor. These cross sections monotonically decline towards stellar energies. The structures in the data that are not explained are possibly due to cluster effects in the nuclear molecule, which are to be included in the present approach.
\end{abstract}

\pacs{24.10.-i, 25.70.Jj, 26.20.Np}

\maketitle

\section{Introduction}
The physics of low-energy nuclear reactions is crucial for understanding the chemical evolution of the Universe \cite{Rolfs}. For instance, $^{12}$C + $^{12}$C fusion at very low energies ($\sim$ 1.5 MeV) plays a key role in stellar carbon burning, whose cross section is commonly determined by extrapolating high-energy fusion data \cite{Spillane,Aguilera,Becker,Cujec,Mazarakis,Patterson}. Direct fusion measurements are very difficult to carry out at very low center-of-mass (c.m.) energies ($\leq$ 3 MeV), with the observed resonant structures making the extrapolation very uncertain \cite{Gasques,Bennett1,Pignatari1}. The reliability of current extrapolation models is also limited by uncertainties associated with the treatment of quantum tunnelling for heavy ions \cite{Back1}. 

The $^{12}$C + $^{12}$C fusion cross sections at very low energies are critical for modelling energy generation and nucleosynthesis during the carbon burning phase of stellar evolution of massive stars ($M \geq 8 M_{\odot}$) \cite{Bennett1,Pignatari1}. These cross sections also determine the ignition conditions for type-Ia supernova explosions \cite{Gasques,Parikh}. Variations of the fusion rate in its traditional range of uncertainty moderately affect nucleosynthesis in the actual type-Ia explosion event \cite{Bravo1}. This situation would change if resonant structures in the low-energy range of the fusion cross sections existed \cite{Spillane}. Such structures have been observed at higher energies and are associated with molecular states \cite{Aguilera,Becker,Cujec,Mazarakis,Patterson}. The possible existence of these states at very low energies can significantly affect nucleosynthesis in type-Ia supernova \cite{Bravo2} as well as superbursts on accreting neutron stars \cite{Steiner}. It is therefore important to go beyond the traditional potential-model approach for averaged cross-sections, to understand the nature of these molecular phenomena and their occurrence at very low energies. 

The fusion of $^{12}$C + $^{12}$C at energies below the Coulomb barrier has been recently addressed with the conventional coupled-channels model \cite{Esbensen,Pierre}. These calculations suggest important effects on the fusion cross section of both the low-lying energy spectrum of $^{24}$Mg and the Hoyle state of $^{12}$C \cite{structure2}. In contrast to experimental observations \cite{Aguilera,Becker,Cujec,Mazarakis,Patterson}, these theoretical fusion excitation curves are smooth, without resonant structures. 
The questions arising here are: what is the origin of the resonant structures in the experimental fusion excitation function? Is this due to a mechanism connected with the physics of the intermediate (nuclear molecule) structure \cite{Greiner}? Why has the conventional coupled-channels model not explained the resonant structures? These important questions are addressed in the present paper.  
 
The key role of intermediate structure in fusion can also be addressed with a novel quantum dynamical model that deals with specific alignments between the $^{12}$C nuclei \cite{Alexis0}. Conclusive results of this model based on wave-packet dynamics \cite{Yabana,Boselli1} are reported in the present paper. The present method directly solves the time-dependent Schr\"odinger equation with a collective Hamiltonian, without the traditional expansion in a basis of energy eigenstates, which is used in the conventional coupled-channels model. Despite this, the numerically calculated total wave function accounts for all the coupled-channel effects. First we present a description of the model and methods, followed by results and a summary.  

\section{Model and Methods}  

\subsection{General aspects of the time-dependent wave-packet method}

The time-dependent wave-packet (TDWP) method involves three steps:
\begin{itemize}
\item[(i)] the definition of the initial wave function $\Psi(t=0)$,
\item[(ii)] the propagation $\Psi(0) \to \Psi(t)$, dictated by the time evolution operator, $\exp (-i \hat{H} t/\hbar )$, where $\hat{H}$ is the total Hamiltonian that is time-independent,
\item[(iii)] after a long propagation time, the calculation of observables (cross sections, spectra, etc) from the time-dependent wave function, $\Psi(t)$.
\end{itemize}

The wave function and the Hamiltonian are represented in a multi-dimensional grid. In this work, these are considered a function of a \emph{few} collective coordinates that include the internuclear distance, thus reducing the complexity of the quantum many-body reaction problem. Moreover, the wave function is \emph{not} expanded in any intrinsic basis (e.g., rotational or vibrational states of the individual nuclei), but it is calculated directly. The \emph{irreversible} process of fusion at small internuclear distances is described with an absorptive potential for fusion. The heavy-ion collision is described in the rotating center-of-mass frame within a nuclear molecular picture \cite{Greiner}.
 
Expressions for the kinetic-energy operator and the collective potential-energy surface, which form the collective Hamiltonian $\hat{H} = \hat{T} + \hat{V}$, are provided in Appendices \ref{APPEX1} and \ref{APPEX2}, respectively. Appendix \ref{APPEX3} describes the time propagator.  

\subsection{The total collective Hamiltonian}
    
\begin{figure}
\centering
\includegraphics[width=8.5cm,clip]{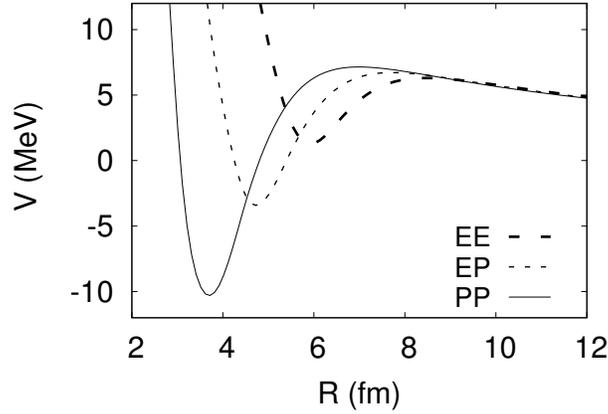}
\caption{Specific cuts in the collective potential-energy landscape of the $^{12}$C + $^{12}$C system as a function of the internuclear distance and three alignments: Equator-Equator (EE), Equator-Pole (EP), Pole-Pole (PP). The EE alignment (dashed line) facilitates the access by tunneling to the potential pockets. All the alignments coexist and compete with each other, the kinetic energy operator driving the system towards either reseparation or fusion in the potential pocket of the PP alignment (solid line) \cite{Alexis1}.}
\label{fig-1}       
\end{figure}

Figure~\ref{fig-1} shows specific cuts in the collective potential-energy landscape of the $^{12}$C + $^{12}$C system as a function of both the internuclear distance and the alignment between the two oblate $^{12}$C nuclei \cite{DEFORM}. The potential curves are presented for fixed orientation of the $^{12}$C-nuclei symmetry axes relative to the internuclear axis, the three axes being coplanar in Fig.~\ref{fig-1}. All the alignments between the $^{12}$C nuclei are included in the dynamical calculations below. The potential energy weakly depends on the angle between crossed symmetry axes of the $^{12}$C nuclei. The overlap between the $^{12}$C nuclei is small at the orientation-dependent potential pockets. The collective potential energy has been calculated using the finite-range liquid-drop model with both universal parameters \cite{FRLDM} and nuclear shapes from a realistic two-center shell model \cite{Alexis1,adt_scheid}. Although shell and pairing corrections to the potential are not included, the volume conservation of the compact dinuclear shapes \cite{Alexis1} guarantees that the effects of nuclear incompressibility on the potential are included. The observed oblate deformation of $^{12}$C \cite{DEFORM} ($\beta_2 = - 0.5$ and moment of inertia $I = 0.67 \, \hbar^2 \textnormal{MeV}^{-1}$ which is derived from the experimental $2^+$ excitation energy [$E_{2^+}= \frac{\hbar^2}{2I}\,2(2+1)=4.44$ MeV]) results in a continuum of Coulomb barriers and potential pockets, which are distributed over a broad range of radii. The lowest barrier (dashed line) favors the initial approach of $^{12}$C nuclei which must re-orientate in order to get trapped in the deepest pocket of the potential (solid line) where fusion occurs \cite{Alexis1}. In transit to fusion, the $^{12}$C + $^{12}$C nuclear molecule can populate quasi-stationary (doorway) states belonging to the shallow potential pockets of the non-axial symmetric configurations. These doorway states may also decay into scattering states, instead of feeding fusion, as the $^{12}$C nuclei largely keep their individuality within the molecule \cite{Alexis1}. The complex motion of the  $^{12}$C + $^{12}$C system through the potential energy landscape is driven by the kinetic energy operator that includes the Coriolis interaction \cite{Poirier} between the total angular momentum of the dinuclear system and the intrinsic angular momentum of the $^{12}$C nuclei. We use an exact expression of the kinetic energy operator \cite{Gatti}. 
           
\subsection{Initial conditions and time propagation}

Having determined the total collective Hamiltonian of the $^{12}$C + $^{12}$C system in terms of the radial coordinate, $R$, and the spherical coordinate angles of the $^{12}$C symmetry axis relative to the internuclear axis, $\theta_i$ and $\phi_i$, the time propagation of an \textit{initial} wave function has been determined using the modified Chebyshev propagator for the evolution operator as described in Appendix \ref{APPEX3}. The initial wave function is determined when the $^{12}$C nuclei are far apart in their ground-states ($j^{\pi} = 0^+$), the radial and the internal coordinates being decoupled:
\begin{equation}
\Psi_0(R,\theta_1,k_1,\theta_2,k_2)\,=\,\chi_0(R) \,\psi_0(\theta_1,k_1,\theta_2,k_2),
\label{eq1}
\end{equation}
where $k_i$ are conjugate momenta of the $\phi_i$ azimuthal angles, so Eq. (\ref{eq1}) is in a mixed representation. Because the radial and internal coordinates are strongly coupled when the $^{12}$C nuclei come together, the product state is only justified asymptotically. $\psi_0(\theta_1,k_1,\theta_2,k_2)$ is the internal symmetrized wave function due to the exchange symmetry of the system:
\begin{eqnarray}
\psi_0(\theta_1,k_1,\theta_2,k_2)=\bigl[ \zeta_{j_1,m_1}(\theta_1,k_1)\zeta_{j_2,m_2}(\theta_2,k_2) \nonumber \\
 + (-1)^J \zeta_{j_2,-m_2}(\theta_1,k_1)\zeta_{j_1,-m_1}(\theta_2,k_2) \bigl]
\nonumber \\
 / \sqrt{2 + 2\,\delta_{j_1,j_2}\delta_{m_1,-m_2}},
\label{eq2}
\end{eqnarray}
where $\zeta_{j,m}(\theta,k) = \sqrt{\frac{(2j+1)(j-m)!}{2\,(j+m)!}}
\,P_j^m (\textnormal{cos}\, \theta) \,\delta_{km}$, $P_j^m$ are associated Legendre functions, and $J$ denotes the total (even) angular momentum. The functions $\zeta_{j,m}(\theta,k)$ describe the individual $^{12}$C nuclei as quantum rigid rotors. In the $^{12}$C ground-state, $j_1 = j_2 = 0$ and $m_1 = m_2 = 0$. The radial component in Eq. (\ref{eq1}), $\chi_0(R)$, is considered a Gaussian wave-packet which contains different translational energies, so an energy projection method is required.

\begin{figure}
\centering
\includegraphics[width=8.5cm,clip]{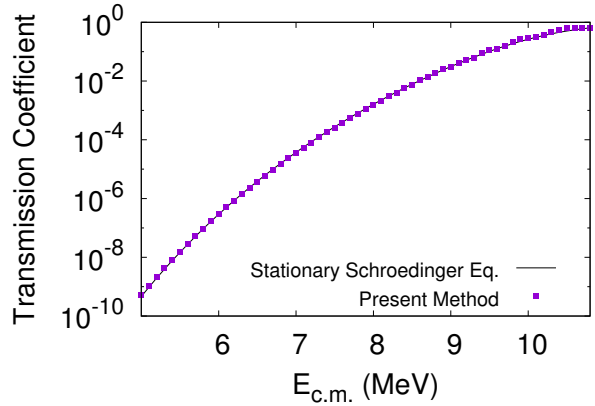}
\caption{(Color online) Transmission-coefficient excitation function for $^{16}$O + $^{16}$O central collisions through the Coulomb barrier of the Broglia-Winther potential \cite{Alexis2}, calculated with the two methods indicated. The barrier height is $\sim 10$ MeV.}
\label{fig-2}       
\end{figure} 

\begin{figure}
\centering
\includegraphics[width=8.5cm,clip]{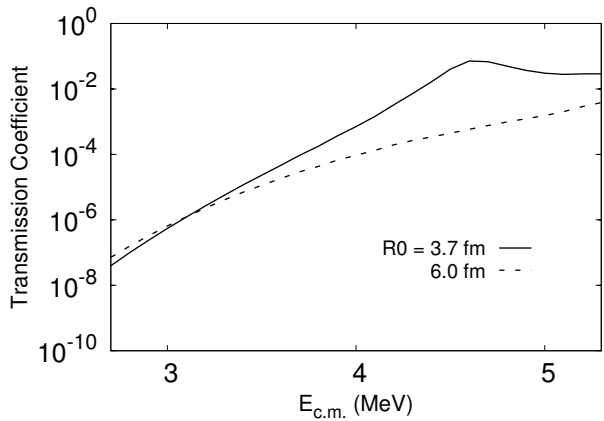}
\caption{The same as in Fig. \ref{fig-2}, but for $^{12}$C + $^{12}$C central collisions. The calculations are performed with the present method using the fusion imaginary potential centered at two different radii as indicated. The resonant structure disappears when this absorption operates around the potential pockets of dinuclear configurations other than the PP configuration in Fig. \ref{fig-1} (comparing the dotted and solid lines).}
\label{fig-3}       
\end{figure} 

\subsection{The energy projection method and transmission coefficients}

The energy-resolved transmission coefficients can be obtained using a window operator \cite{Schafer}. The key idea is to calculate the energy spectrum of the initial and final wave functions, the initial spectrum corresponding to a Gaussian distribution centered at the mean energy $E_0$. The energy spectrum is $\mathcal{P}(E_k) \,=\, \langle \Psi | \hat{\Delta} |\Psi \rangle $, where $\hat{\Delta}$ is the window operator \cite{Schafer}:
\begin{equation}
\hat{\Delta}(E_k,n,\epsilon)\, \equiv \, \frac{\epsilon^{2^n}}  
{(\hat{\mathcal{H}}\, - \, E_k)^{2^n}\, + \, \epsilon^{2^n}}, 
\label{eq5}
\end{equation}      
$\hat{\mathcal{H}}$ is the system asymptotic Hamiltonian when the $^{12}$C nuclei are well separated, and $n$ determines the shape of the window function. As $n$ is increased, this shape rapidly becomes rectangular with very little overlap between adjacent energy bins with centroid $E_k$, the bin width remaining constant at $2\epsilon$ \cite{Schafer}. The spectrum is constructed for a set of $E_k$ where $E_{k+1}=E_k + 2\epsilon$. Thus, scattering information over a range of incident energies can be extracted from a time-dependent numerical wave function. In this work, $n=2$ and $\epsilon=50$ keV \cite{Schafer}. Solving two successive linear equations for the vector $|\chi \rangle$:
\begin{equation}
(\hat{\mathcal{H}}\, - \, E_k \, + \, \sqrt{i}\,\epsilon)(\hat{\mathcal{H}}\, - \, E_k \, - \, \sqrt{i}\,\epsilon)\, |\chi \rangle \,=\, |\Psi \rangle,
\label{eq6}
\end{equation}
yields $\mathcal{P}(E_k) \,=\, \epsilon^4 \, \langle \chi |\chi \rangle $.  The state $|\chi \rangle$ represents the scattering state with a definite energy $E_k$. Eq. (\ref{eq6}) is solved for both the initial and final wave functions.

\begin{figure}
\centering
\includegraphics[width=8.5cm,clip]{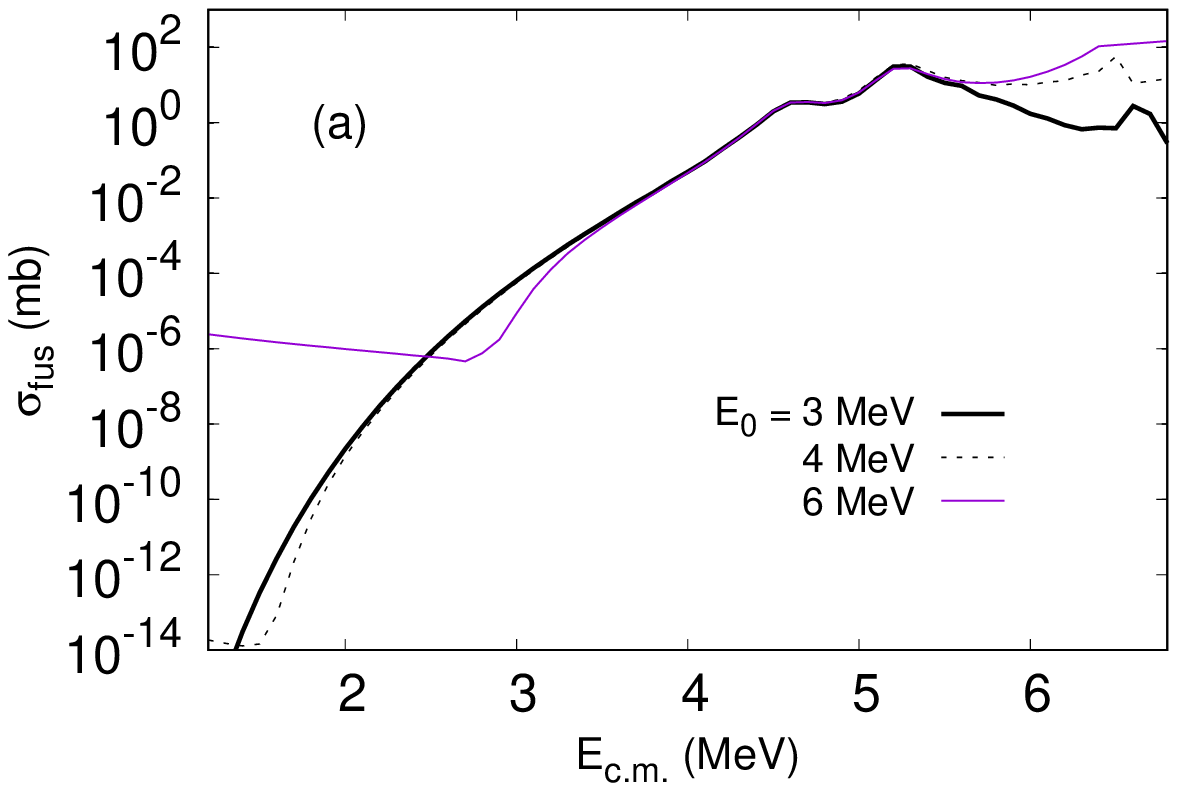} \\
\includegraphics[width=8.5cm,clip]{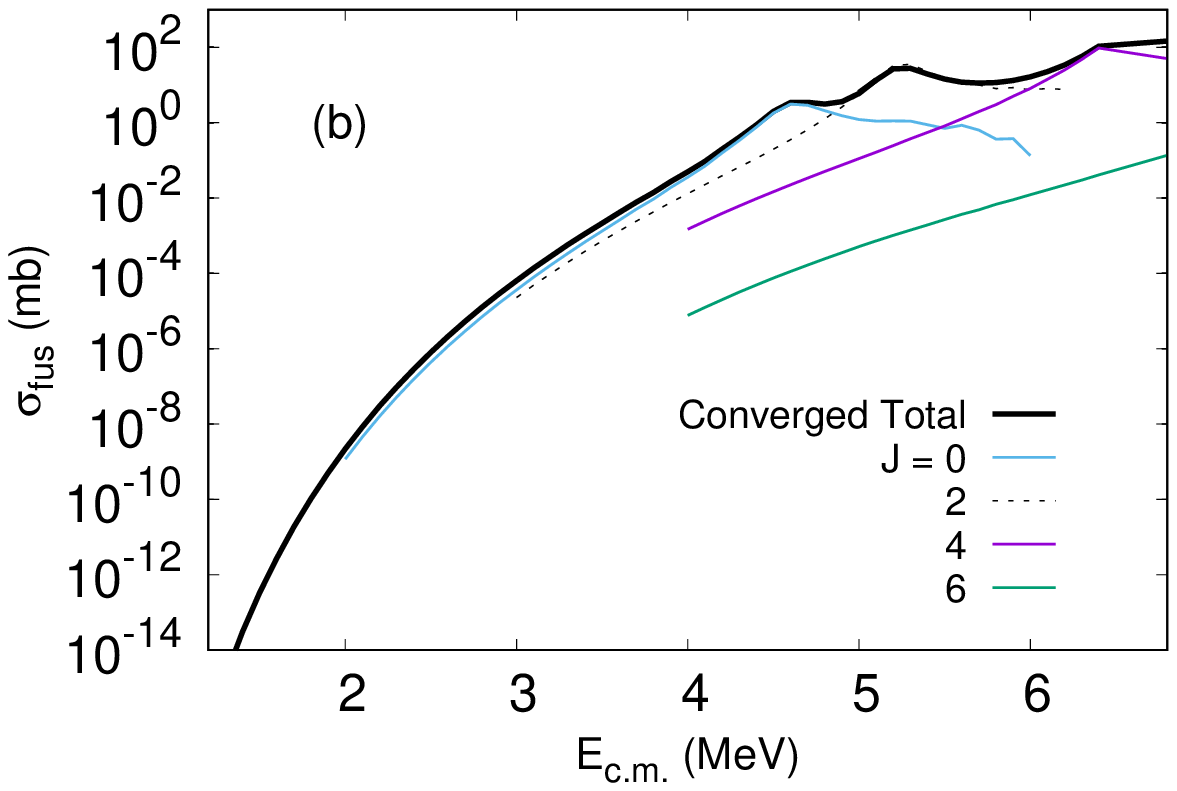}
\caption{(Color online) (a) Energy-resolved fusion cross sections for different values of the mean energy $E_0$ of the initial wave function. The energy components far from $E_0$ have small amplitudes, and the associated cross sections are very inaccurate. Only the overlapping parts of 
these excitation curves determine the physical, converged fusion excitation function. (b) Angular momentum decomposition of the converged fusion excitation function that shows some bumps originating from specific partial waves.}
\label{fig-4}       
\end{figure}

\begin{figure}
\centering
\includegraphics[width=10.0cm,clip]{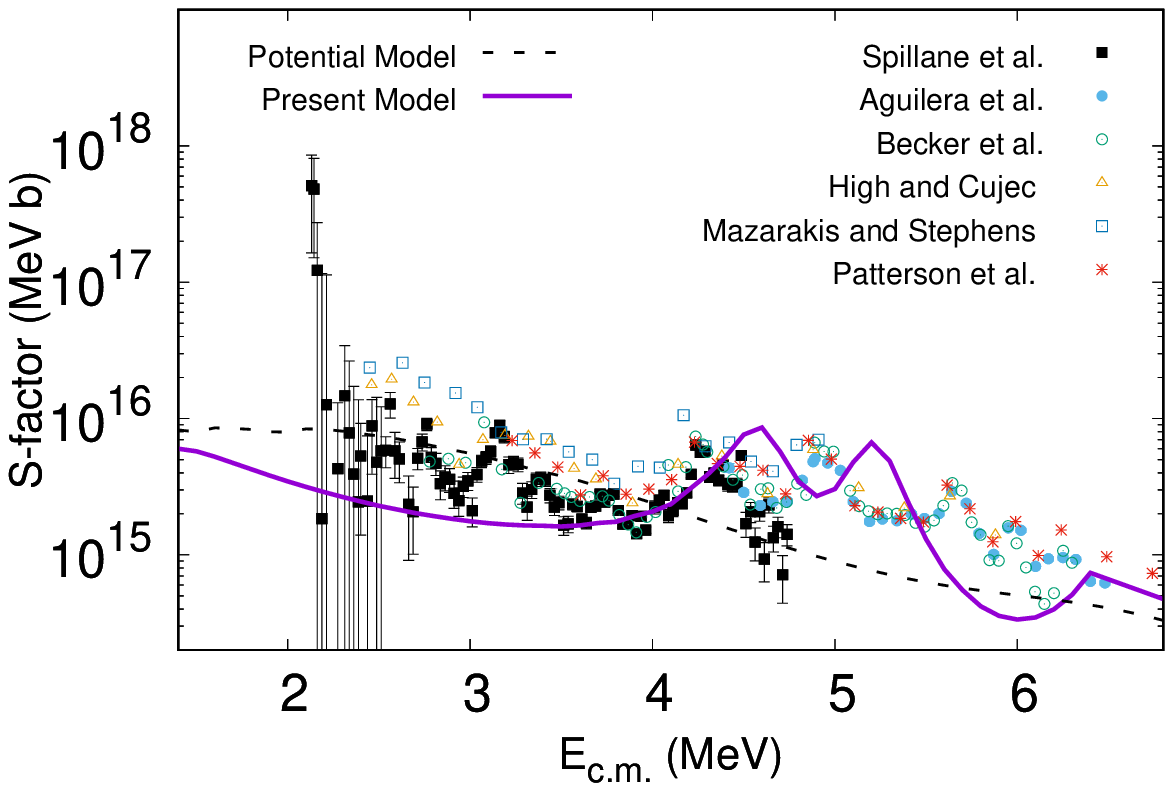}
\caption{(Color online) The astrophysical S-factor excitation function for $^{12}$C + $^{12}$C. Measurements \cite{Spillane,Aguilera,Becker,Cujec,Mazarakis,Patterson} (symbols) are compared to model calculations (dashed and solid lines), indicating that molecular structure and fusion are interconnected.}
\label{fig-5}       
\end{figure}

The transmission coefficients are obtained from:
\begin{equation}
\mathcal{T}(E_k)\, = \, \frac{-(8/\hbar v_k) \, \epsilon^4 \, \langle \chi | \textnormal{Im}(\hat{W}) |\chi \rangle}{\mathcal{P}^{\, initial}(E_k)},
\label{eq7}
\end{equation}
where  $v_k = \sqrt{2E_k/\mu}$ is the asymptotic relative velocity, $\mu$ is the reduced mass, and $\textnormal{Im}(\hat{W}) < 0$ denotes the strong, imaginary Woods-Saxon potential centered at the minimum of the PP potential pocket in Fig \ref{fig-1} (solid line), which operates very weakly at the potential pockets of the non-axial symmetric configurations. The strong repulsive core of the real potentials for non-axial symmetric dinuclear configurations hinders the effect of the imaginary fusion potential on the potential resonances formed in the corresponding real potential pockets. 

\subsection{The role of the fusion absorption}

The functional form of the imaginary fusion potential is $W = W_0/[1+\exp((R-R^{PP}_{min})/a_{0w})]$, where the strength $W_0 = -50$ MeV, the diffuseness $a_{0w} = 0.2$ fm, and $R^{PP}_{min}= 3.7$ fm. This imaginary potential is usually employed in the coupled-channels model to simulate fusion and is equivalent to the use of the ingoing-wave boundary condition (IWBC) \cite{Boselli1,AlexisThompson1}. 
As a simple example, Fig. \ref{fig-2} shows the transmission-coefficient excitation function for $^{16}$O + $^{16}$O central collisions, which are determined by two methods: solving the stationary Schr\"{o}dinger equation with IWBC \cite{Alexis2} (solid line) and employing Eq. (\ref{eq7}) (symbols). The good agreement between the two methods demonstrates the reliability of Eq. \ref{eq7}. 

Using the present method, Fig. \ref{fig-3} shows the effective transmission coefficient for head-on collisions of $^{12}$C + $^{12}$C through the Coulomb barriers presented in Fig. \ref{fig-1}. The calculations are carried out using the fusion imaginary potential centered either at $R^{PP}_{min}$ (solid line) or around the potential pockets for non-axial symmetric dinuclear configurations (dotted line), demonstrating the crucial role of this fusion absorption in the appearance of resonant structures in the fusion excitation function. The results shown in Fig. \ref{fig-3} do not change if the strength of the absorption is reduced by a factor of two. Fig. \ref{fig-3} demonstrates how increasing the range of the imaginary fusion potential affects a potential resonance for $J =0$. As expected, the resonant structure of the transmission coefficient for fusion dissapears.  

\subsection{Fusion cross sections and the astrophysical S-factor}

The fusion cross section, $\sigma_{fus}(E)$, is calculated taking into account the identity of the interacting nuclei and the parity of the radial wave function (only even partial waves $J$ are included), i.e., $\sigma_{fus}(E) = \pi \hbar^2 /(\mu E) \sum_J (2J + 1)\mathcal{T}_J(E)$, where $E$ is the incident c.m. energy and $\mathcal{T}_J$ is the partial transmission coefficient. The S-factor is $S(E) = \sigma_{fus}(E) E \exp (2 \pi \eta)$, where the Sommerfeld parameter $\eta = (\mu/2)^{1/2}Z_1 Z_2 e^2 / (\hbar E^{1/2})$ and $Z_i=6$ is the $^{12}$C charge number.

\section{Numerical details and Results}
The model calculations are performed on a five-dimensional grid, i.e., a Fourier radial grid ($R=0-1000$ fm) with $2048$ evenly spaced points \cite{Kosloff2}, and for the angular variables, ($\theta_1,k_1$) and ($\theta_2,k_2$), a grid based on the extended Legendre discrete-variable representation (KLeg-DVR) method \cite{Meyer1}. The KLeg-DVR grid-size is determined by the values of the $^{12}$C intrinsic $j_{max}$ and $k_{max}$ \cite{Meyer1}, which are set as $4$, and this grid also supports odd $j$ values. The initial wave-packet was centered at $R_0 = 400$ fm, with width $\sigma = 10$ fm, and was boosted toward the collective potential-energy landscape with the appropriate kinetic energy for the mean energy $E_0$ required. In this work, $\Delta t = 10^{-22}$ s, and in the absence of the imaginary potential the norm of the wave function is preserved with an accuracy of $\sim 10^{-14}$. All the parameters for the grid and the initial wave-packet guarantee the convergence of the calculated fusion cross sections \cite{Boselli1}.   

The energy-resolved fusion cross sections are provided by a few wave-packet propagations with $E_0 = 3, 4$ and $6$ MeV and total angular momenta up to $J=6 \hbar$, as shown in Fig. \ref{fig-4}. Fig. \ref{fig-4}(a) shows the convergence relative to $E_0$ of the energy-resolved fusion excitation function, while Fig.~\ref{fig-4}(b) presents the angular momentum decomposition of the converged fusion excitation curve. It can be seen that the converged fusion curve shows, at energies slightly below the nominal Coulomb barrier ($\sim 6.5$ MeV), three maxima caused by specific partial waves. The total angular momenta up to $J=4 \hbar$ determine the sub-Coulomb fusion cross sections which monotonically decline towards stellar energies. 

\begin{figure}
\centering
\includegraphics[width=10.0cm,clip]{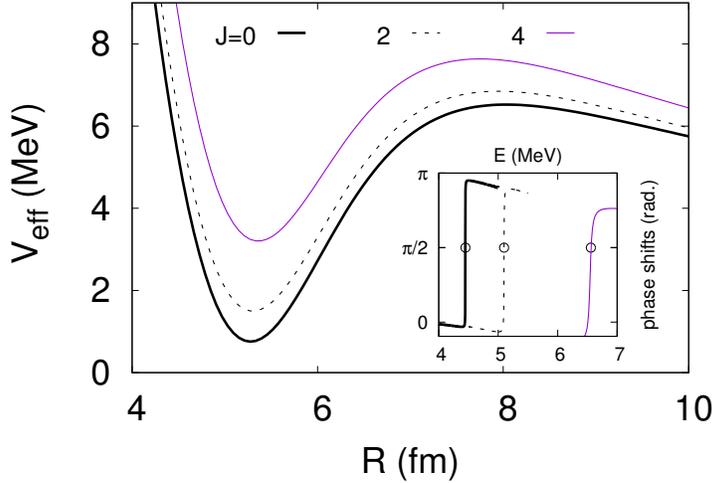}
\caption{(Color online) The effective real potentials for non-axial 
symmetric molecular configurations and the scattering phase-shift analysis (plot inserted) for specific partial waves. The occupation of these potential resonances (circles in plot inserted) causes the structures in the S-factor excitation function in Fig. \ref{fig-5}.}
\label{fig-6}       
\end{figure}

The sub-Coulomb S-factor excitation function for $^{12}$C + $^{12}$C is presented in Fig. \ref{fig-5} which shows key features: 
\begin{enumerate}
\item  The observed resonant structures in the 4-6.5 MeV energy window are qualitatively reproduced by the present model calculations (solid line). The positions of the theoretical maxima are shifted by $\sim 0.3$ MeV with respect to the experimental maxima, and that position is determined by the features (depth and curvature) of the potential pockets for non-axial symmetric configurations. Those pockets support intermediate molecular states. Fig. \ref{fig-6} shows the first quasi-bound molecular states in the effective real potentials for specific partial waves. Although these effective potentials are not used in solving the time-dependent Schr\"odinger equation, they are useful for understanding the formation of molecular resonance states. These effective potentials are determined by folding the potential energy (including the centrifugal energy) of non-axial symmetric configurations with the probability density of the initial wave-function in Eq. (\ref{eq2}). A scattering phase-shift analysis (plot inserted in Fig. \ref{fig-6}) provides values of $4.45$ MeV ($J=0$), $5.10$ MeV ($J=2$) and $6.56$ MeV ($J=4$) which are consistent with the positions of the theoretical maxima in Figs. \ref{fig-4}(b) and \ref{fig-5}. Unlike the conventional coupled-channels method, the present method keeps these molecular resonance states visible through the treatment of absorption as demonstrated in Fig. \ref{fig-3}.  
\item  At deep sub-barrier energies ($E < 4$ MeV), the S-factor of the present model (solid line) is smooth and slightly underestimates the experimental data. Some experimental data \cite{Spillane,Becker,Cujec} show a resonant structure around $3.1$ MeV which is not explained by the present model. There are great variations among data sets. More accurate measurements are required at the astrophysically important energy region, $E \leq 3$ MeV, which are challenging and are being pursued currently \cite{Exp1-12C,Exp2-12C,Exp3-12C}. The predicted smooth S-factor around the Gamow peak ($\sim 1.5$ MeV) would moderately change the present abundance distributions in type-Ia supernova \cite{Bravo2}, and would require a closer look at the hydrodynamics of superbursts to enhance the energy output \cite{Steiner}.  
\item For comparison, a potential-model calculation has been carried out. It is based on solving the stationary Schr\"{o}dinger equation with IWBC for each partial wave and alignment between the $^{12}$C nuclei, and averaging the partial transmission coefficients over all the alignments. The S-factor excitation curve (dashed line) is smooth, like those resulting from the coupled-channels model in Refs. \cite{Esbensen,Pierre}. Comparing the dashed to the solid line, it is observed that the effects of the intermediate structure on fusion are crucial.   
\end{enumerate}

The degree of agreement between the present model calculations and the experimental data in Fig. \ref{fig-5} may be improved by (i) using a more complete potential-energy landscape that includes both shell and pairing corrections which may modify the features of the potential pockets, and (ii) releasing and treating explicitly the alpha-particle degrees of freedom which should lead to the fragmentation of the obtained resonant structures \cite{Vogt}. We have included the effects of the Coriolis interaction in the present calculations, and these effects are very weak for the relevant partial waves ($< 1 \%$).

\begin{figure}
\centering
\includegraphics[width=10.0cm,clip]{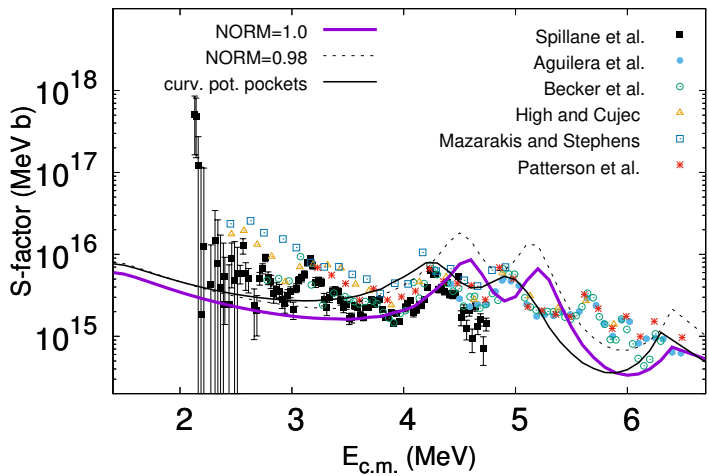}
\caption{(Color online) The same as in Fig. \ref{fig-5}, but the present model calculations are shown for (i) two global factors that multiply the collective potential-energy landscape in Fig. \ref{fig-1}, and (ii) a reduction by 15$\%$ of the curvature of the potential pockets. The latter greatly improves the location of the predicted resonant structures (thin solid line).}
\label{fig-7}       
\end{figure}

Multiplying the collective potential-energy landscape in Fig. \ref{fig-1} by a global factor of $0.98$ slightly improves the position of the calculated resonant structures relative to those observed in the 4-6.5 MeV energy window by $\sim 0.1$ MeV, as shown in Fig. \ref{fig-7}. There remains a mismatch of $\sim 0.2$ MeV that cannot be removed by the potential renormalization. However, reducing the curvature of the potential pockets in Fig. \ref{fig-1} by 15$\%$ significantly improves the location of the predicted resonant structures (thin solid line). Some dynamical effects, that are not yet included in the model, may yield additional resonant structures in the S-factor. For instance, we use an oblately deformed rigid-rotor model for the $^{12}$C nuclei, but in reality they may also be vibrating \cite{structure2}. Both the vibrations and the rotation-vibration interaction affect the resonance energies of the molecular states \cite{Scheid2}. Cluster effects in the nuclear molecule (e.g., $^{20}$Ne + alpha and $^{23}$Na + p) can also be very important. 

\subsection{Coupled-channels calculations and the TDWP method}   

In contrast to the simplified coupled-channels model of Ref. \cite{Abe} that uses a weak absorption, the sophisticated coupled-channels calculations of Refs. \cite{Esbensen,Pierre}, which are also conventional but use strong absorption, do \textit{not} produce any resonant structure. This indicates the importance of treating explicitly the dynamics of the intermediate (nuclear molecule) configurations inside the radius of the nominal Coulomb barrier. Weak absorption may allow that kind of treatment \cite{Abe}, which also requires the inclusion of highly excited states in the individual $^{12}$C nuclei, well beyond their first $2^+$ excited states \cite{Rowley2016}. The conventional coupled-channels model \cite{Esbensen,Pierre, Abe, Rowley2016} does not address \textit{specific} alignments between the $^{12}$C nuclei, but uses an \textit{average} over all the alignments, i.e., there is an integration over orientation angles in the coupling potentials. Unconventional coupled-channels calculations could deal with \textit{specific} dinuclear configurations \cite{Maass}. The present model is based on the TDWP method which is novel in this field, allowing one to address more physical details about the dynamics of compact dinuclear configurations: (i) the implicit orientation-dependent fusion absorption, justified with the two-center shell model \cite{Alexis1}, provides a weak absorption for most intermediate (nuclear molecule) configurations, and (ii) these configurations are treated naturally. The fact that (i) is absent from the conventional coupled-channels model is the main reason why the coupled-channels model has not reproduced resonant structures but average trends of the sub-barrier fusion excitation function for $^{12}$C + $^{12}$C.   

\section{Summary}
The present quantum dynamical model indicates that molecular structure and fusion are closely connected in the $^{12}$C + $^{12}$C system, suggesting that the fusion excitation function monotonically declines towards stellar energies. The fusion imaginary potential for specific alignments between the $^{12}$C nuclei is crucial for understanding the appearance of resonances in the fusion cross section. In contrast to other commonly used methods, such as the potential model and the conventional coupled-channels approach, these new calculations reveal three resonant structures in the S-factor for fusion. The structures correlate with similar structures in the data. The structures in the data that are not explained are possibly due to cluster effects in the nuclear molecule, which are to be included in the new approach. This method is a suitable tool for extrapolating the cross section predictions towards stellar energies. 

\acknowledgments
The support from the STFC grant (ST/P00671X/1) is acknowledged. We thank Leandro Gasques, Zsolt Podolyak and Phil Walker for constructive comments, and Jeff Tostevin for the scattering phase-shift analysis shown in Fig. \ref{fig-6}.

\appendix
\section{Collective Kinetic-Energy Operator}
\label{APPEX1}

The collision of two arbitrarily oriented, deformed nuclei can be described in the rotating center-of-mass frame with five collective coordinates: the internuclear distance $R$, and the ($\theta_1, \phi_1$) and ($\theta_2, \phi_2$) spherical angles of the nuclei symmetry axis. Using a mixed representation for the internal angular-momentum operators $\hat{j}_i$ (i.e., coordinate representation for the polar angles $\theta_i$ and a momentum representation, $k_i$, replacing the azimuthal angles $\phi_i$), the exact kinetic-energy operator in the rotating frame is \cite{Gatti}:
\begin{eqnarray}
\frac{2\hat{T}}{\hbar^2} &=& -\frac{1}{\mu}\frac{\partial^2}{\partial R^2}\,+\,
\bigl( \frac{1}{I_1}+\frac{1}{\mu R^2} \bigl) \hat{j}^2_1
\,+\, \bigl( \frac{1}{I_2}+\frac{1}{\mu R^2} \bigl) \hat{j}^2_2 \nonumber \\
&& \,+\, \frac{1}{\mu R^2} \bigl[ \hat{j}_{1,+}\hat{j}_{2,-} \,+\,
\hat{j}_{1,-}\hat{j}_{2,+} \,+\, J(J+1) \nonumber \\ 
&& \,-\,2k^2_1 \,-\, 2k_1k_2 \,-\, 2k^2_2 \bigl] \,-\, \frac{C_{+}(J,K)}{\mu R^2} \bigl( \hat{j}_{1,+} \,+\, \hat{j}_{2,+} \bigl) \nonumber \\
&&\,-\, \frac{C_{-}(J,K)}{\mu R^2} \bigl( \hat{j}_{1,-} \,+\, \hat{j}_{2,-} \bigl),
\label{C1}
\end{eqnarray}
where $\mu$ is the reduced mass for the radial motion, $I_i$ is the rotational inertia of the nuclei, $J$ is the total angular momentum with projection $K=k_1 + k_2$, $C_{\pm}(J,K) = \sqrt{J(J+1) \,-\, K(K \pm 1)}$, and $\hat{j}^2_i \,=\, -\frac{1}{\textnormal{sin}\,\, \theta_i}\frac{\partial}{\partial \, \theta_i}\textnormal{sin}\,\, \theta_i \frac{\partial}{\partial \, \theta_i} \,+\, \frac{k^2_i}
{\textnormal{sin}^2 \,\, \theta_i},$ and $\hat{j}_{i,\pm} \,=\, \pm \frac{\partial}{\partial \, \theta_i} \,-\, k_i \,
\textnormal{cot} \,\, \theta_i$. When the $\hat{j}_{i,\pm}$ operators act on the $k_i$ component of the wave function, the outcomes emerge in its $k_i \, \pm \, 1$ component. The last two terms in Eq. (\ref{C1}) describe the Coriolis interaction that changes the $K$ quantum number. The $^{12}$C nuclei keep their individuality within most dinuclear configurations, as demonstrated in Ref. \cite{Alexis1}.

\section{Collective Potential-Energy Surface}
\label{APPEX2}

Macroscopic nuclear and Coulomb energies of a dinuclear system can be determined within the finite-range liquid-drop model \cite{FRLDM}. 
The nuclear component reads as:
\begin{eqnarray}
E_n &=& - \, \frac{c_s}{8\pi^2 r_0^2} \, \oint_S \oint_{S'} \, f(\sigma) \, (\vec{\sigma} d\vec{S})(\vec{\sigma} d\vec{S'}) \nonumber \\
&=& - \, \frac{c_s}{8\pi^2 r_0^2} \oint_T \oint_{T'} \, f(\sigma) \, \vec{\sigma} \, (\frac{\partial \vec{r}}{\partial \phi} 
\times \frac{\partial \vec{r}}{\partial z}) \, d\phi dz  \nonumber \\
&& \ast \, \vec{\sigma} \, (\frac{\partial \vec{r'}}{\partial \phi'} 
\times \frac{\partial \vec{r'}}{\partial z'}) \, d\phi' dz',
\label{B1} 
\end{eqnarray}
where $f(\sigma)=\{2-[(\frac{\sigma}{a})^2 + 2\frac{\sigma}{a} + 2]e^{-\frac{\sigma}{a}}\}\sigma^{-4}$ and $\vec{\sigma} =\vec{r}-\vec{r'}$.

The vectors $\vec{r}\equiv[P(z,\phi)\textnormal{cos}\phi,P(z,\phi)\textnormal{sin}\phi,z]$ and 
$\vec{r'}\equiv[P(z',\phi')\textnormal{cos}\phi',P(z',\phi')\textnormal{sin}\phi',z']$ determine the position of 
the nuclear surface elements $d\vec{S}$ and $d\vec{S'}$, respectively. These are characterized by cylindrical 
coordinates [$z_{min} \leq z \leq z_{max}$ and $0\leq \phi \leq 2\pi$ define the $T$ integration region] where $P(z,\phi)$ denotes the distance 
from the surface elements to the $z$-axis that contains the origin of the coordinate system. $P(z,\phi)$ is determined using the volume-conserving nuclear shapes of the two-center shell model \cite{Alexis1,adt_scheid}.

In terms of $P(z,\phi)$, $\sigma$ and $E_n$ in Eq. (\ref{B1}) are:
\begin{eqnarray}
\sigma^2 &=& P^2(z,\phi) + P^2(z',\phi') - 2 \,P(z,\phi)P(z',\phi') \nonumber \\
&& \ast \, \textnormal{cos}(\phi - \phi') + (z-z')^2,
\label{B2}
\end{eqnarray}
\begin{eqnarray}
E_n &=& - \, \frac{c_s}{8\pi^2 r_0^2} \, \oint_T \oint_{T'} \, f(\sigma) \Big\{ P(z,\phi) \, \Big[ P(z,\phi) \nonumber \\ 
&& - \, P(z',\phi') \, \textnormal{cos}(\phi - \phi') - \frac{\partial P(z,\phi)}{\partial z} \,(z-z') \Big]  \nonumber \\
&& - \, \frac{\partial P(z,\phi)}{\partial \phi} P(z',\phi') \, \textnormal{sin} (\phi - \phi') \Big\} \nonumber \\
&& \ast \, \Big\{ P(z',\phi') \, \Big[ P(z',\phi') - P(z,\phi) \, \textnormal{cos}(\phi - \phi') \nonumber \\
&& + \, \frac{\partial P(z',\phi')}{\partial z'} \,(z-z') \Big]  \nonumber \\
&& + \, \frac{\partial P(z',\phi')}{\partial \phi'} P(z,\phi) \, \textnormal{sin} (\phi - \phi') \Big\} dz dz' d\phi d\phi'. \nonumber \\
\label{B3} 
\end{eqnarray}

For axial-symmetric nuclear shapes, $P=P(z)$, Eq. (\ref{B3}) yields the Krappe-Nix-Sierk formula \cite{KNS}. 

The Coulomb energy reads as:
\begin{equation}
E_C = - \, \frac{\rho_0^2}{12} \, \oint_S \oint_{S'} \, \sigma^{-1} \, (\vec{\sigma} d\vec{S})(\vec{\sigma} d\vec{S'}),
\label{B4}
\end{equation}
where $\rho_0=Ze(4\pi r_0^3 A/3)^{-1}$ is a constant charge density, and the integrals in Eq. (\ref{B4}) are determined 
like in Eq. (\ref{B1}).

Eqs. (\ref{B1}) and (\ref{B4}) correspond to a uniform sharp-surface distribution of given shape. Ref. \cite{FRLDM1} provides expressions 
for an arbitrarily shaped diffuse-surface nuclear density distribution, whose diffuseness correction to the above nuclear and Coulomb energies 
is also included in the present work. 

The above formulae provide total self-energies. The total collective potential energy is $V = E_n + E_C$, whose interaction component is determined by 
substracting the total self-energy of the two individual nuclei.   
When the two interacting nuclei do not overlap with each other \cite{FRLDM}, the constant parameters are $c_s = [c_s(1) c_s(2)]^{1/2}$, $r_0^2 = r_{01} r_{02}$, 
$\rho_0^2 = \rho_{01} \rho_{02}$, where $c_s(i) = a_s[1-\kappa_s(\frac{N_i - Z_i}{N_i + Z_i})^2]$. In this work, the constants of Ref. \cite{FRLDM} are used, 
i.e, $a_s = 21.13$ MeV, $\kappa_s = 2.30$, $e^2 = 1.4399764$ MeV fm, $a=0.68$ fm, and $r_{01}=r_{02}=r_0 = 1.16$ fm.

\section{Modified Chebyshev Propagator}
\label{APPEX3}

The formal solution of the time-dependent Schr\"odinger equation at $t+\Delta t$ is
\begin{equation}
\Psi(t+ \Delta t) = \exp \Big(-i \frac{\hat{H}\,\Delta t}{\hbar} \Big) \Psi(t).
\label{A0}
\end{equation}

The time evolution operator is represented as a convergent series of polynomials 
$Q_n$ \cite{Mandelshtam}:
\begin{equation}
\exp \Big(-i \frac{\hat{H}\,\Delta t}{\hbar} \Big) \, \approx \, \sum_{n} \, a_n \, Q_n (\hat{H}_{norm}). 
\label{A1} 
\end{equation}
In Eq. (\ref{A1}), the time-independent Hamiltonian is renormalized so that its spectral range is within the interval [-1,1], 
the domain of the polynomials, by defining
\begin{equation}
\hat{H}_{norm} = \frac{(\bar{H} \, \hat{1} - \hat{H})}{\Delta{H}},
\label{A2}
\end{equation}
where $\bar{H} = (\lambda_{max} + \lambda_{min})/2$, $\Delta{H} = (\lambda_{max} - \lambda_{min})/2$, $\lambda_{max}$ and $\lambda_{min}$ are respectively the largest and smallest eigenvalues in the spectrum of $\hat{H}$ supported by the grid, and $\hat{1}$ denotes the identity operator. The expansion coefficients in (\ref{A1}) read as
\begin{equation}
a_n = i^n (2-\delta_{n0})\exp ( -i \frac{\bar{H} \, \Delta t}{\hbar} ) \, 
J_n ( \frac{\Delta H \, \Delta t}{\hbar} ),
\label{A3} 
\end{equation}
where $J_n$ are Bessel functions of the first kind. Since $J_n (x)$ exponentially goes to zero with increasing $n$ for $n > x$, 
the expansion (\ref{A1}) converges exponentially for $n > \Delta H \Delta t / \hbar$. 
With a suitable approximation for the spectral range of the Hamiltonian, the expansion (\ref{A1}) numerically represents the time evolution operator.

This representation of the time evolution operator requires the action of $Q_n(\hat{H}_{norm})$ on the wave function $\Psi(t)$. The $Q_n$ polynomials obey the 
recurrence relations \cite{Mandelshtam}:

\begin{eqnarray}
e^{-\hat{\gamma}}Q_{n-1}(\hat{H}_{norm}) + e^{\hat{\gamma}}Q_{n+1}(\hat{H}_{norm}) 
\nonumber \\
- 2\hat{H}_{norm}Q_n(\hat{H}_{norm}) = 0,
\label{A4}
\end{eqnarray} 
with the initial conditions $Q_0(\hat{H}_{norm})=\hat{1}$ and 
$Q_1(\hat{H}_{norm})=e^{-\hat{\gamma}}\hat{H}_{norm}$. Here, $\hat{\gamma}$ is 
an operator related to the absorbing optical potential $\hat{W}$ which 
can be written as \cite{Mandelshtam}:
\begin{equation}
\hat{W} = \Delta H \, [ \textnormal{cos} \,\, \xi \,\, (1 - \textnormal{cosh} \,\, \hat{\gamma})\,\, 
-i \,\, \textnormal{sin} \,\, \xi \,\, \textnormal{sinh} \,\, \hat{\gamma}],
\label{A5}
\end{equation}
where $\xi = \textnormal{arcos}\,(\frac{E-\bar{H}}{\Delta H})$, and $E$ denotes the collision energy. 
If there is no absorption ($\gamma=0$ implies $W=0$), the $Q_n$ polynomials in expression (\ref{A4}) will be the Chebyshev 
polynomials and the expansion (\ref{A1}) will correspond to the Chebyshev propagator \cite{Kosloff2}. Chapter 11 in Ref. \cite{Tannor} 
provides a survey of techniques for solving the time-dependent Schr\"odinger equation, which are distinguished by the numerical 
implementation of the time evolution operator (\ref{A0}).

\end{document}